\documentclass[sigconf,acmtog,nonacm=true]{acmart}
		\usepackage[ruled]{algorithm2e}
		\usepackage{amsmath,bm,dsfont,amsfonts,enumitem}
		\setcopyright{none}
		\copyrightyear{2022}
		\acmYear{2022}
		\acmDOI{}
		\acmPrice{}
		\acmISBN{}
		\acmBooktitle{}
		
		\setlist[itemize]{leftmargin=*}

		\makeatletter
		\newcommand{\removelatexerror}{\let\@latex@error\@gobble}
		\makeatother
		\DeclareMathOperator*{\argmin}{arg\,min}

\begin{document}
	
	\title{DMS, AE, DAA: methods and applications of adaptive time series model selection, ensemble, and financial evaluation}
	
	\subtitle{Date: 5 July 2022, 	Most Recent Version: \href{https://drive.google.com/drive/folders/1YPLWWzZgBegg1h1O-NL27MWyV94V2jzZ?usp=sharing}{Click Here}	}
	
	\author{Parley Ruogu Yang}
	
	\affiliation{%
		\institution{University of Cambridge}
		\country{UK}
	}
	
\email{ry266@cam.ac.uk}

	\author{Ryan Lucas}
	\affiliation{%
		\institution{Massachusetts Institute of Technology}
		\country{USA}
	}
	
	\begin{teaserfigure}
	\centering
	\includegraphics[width=\linewidth]{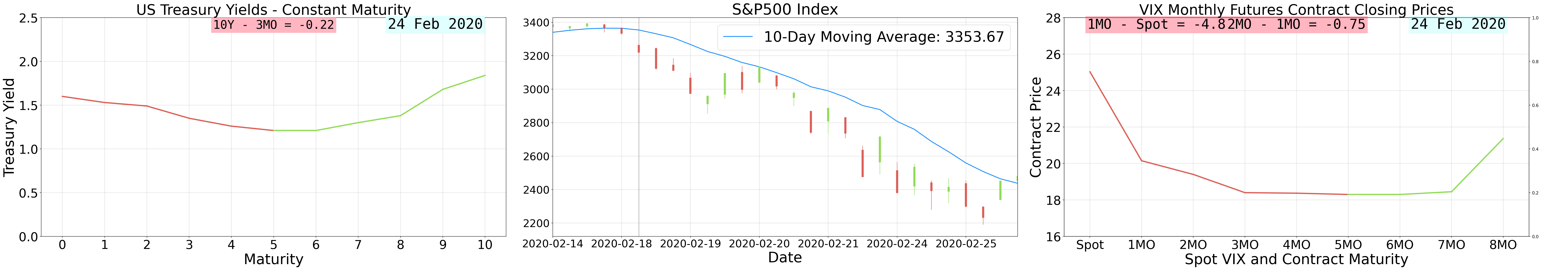}
	\caption{The Yield Curve (left) and VIX Curve (right) alongside the price of the S\&P 500 index (centre) during the pandemic-induced market contraction of 2020. Inversions in these curves preceded the collapse of the index}
	\label{fig0}
	\end{teaserfigure}
\begin{abstract}
	\textbf{Abstract} \\
	We introduce three adaptive time series learning methods, called Dynamic Model Selection (DMS), Adaptive Ensemble (AE), and Dynamic Asset Allocation (DAA). The methods respectively handle model selection, ensembling, and contextual evaluation in financial time series. Empirically, we use the methods to forecast the returns of four key indices in the US market, incorporating information from the VIX and Yield curves. We present financial applications of the learning results, including fully-automated portfolios and dynamic hedging strategies. The strategies strongly outperform long-only benchmarks over our testing period, spanning from Q4 2015 to the end of 2021. The key outputs of the learning methods are interpreted during the 2020 market crash. \\
	
	\noindent \textbf{Acknowledgement}: 
	We thank Piotr Fryzlewicz for offering feedbacks on the drafts,	Camilla Schelpe for discussions during the work, and the Oxford SMLFin Seminar for giving us the opportunity to discuss an earlier version of this work. \\
	\noindent \textbf{Software}: Computing implementations are done in Python, 
and	 are available at \url{https://github.com/parleyyang/DMS-AE-DAA}.

	\end{abstract}
	
	\maketitle

\section{Introduction}
	It is not wise to apply generically-developed machine learning methods in a time series environment. This is because of the unique nature of time series, where observations are not i.i.d., but rather have an ordered dependence structure and potential change-points. The distribution of time series data at a certain time $t$ may differ massively from some time far in the past. By adapting to past forecasting errors in a time series context, Dynamic Model Selection (DMS) and Adaptive Ensemble (AE) take a rolling view of model selection and ensembling respectively. This differs from typical cross-validation procedures, which attempt to find a single best model for testing, rather than provide ongoing selections. 
	
	The DMS and AE approaches still face two problems: how to tune the hyperparameters over time? And how to compare models that forecast different variables or horizons? For instance, some assets may be easier to forecast than others and shorter forecast horizons may induce lower errors than longer ones. This motivates the third algorithm, Dynamic Asset Allocation (DAA), which is a dynamic evaluation mechanism based on Sharpe Ratio (SR). In this paper, we apply all three algorithms to create automated portfolio strategies and demonstrate their strong performance against benchmarks in both return and risk metrics. We also demonstrate the interpretability of our methods by obtaining the relevant information produced by the algorithms, and further in conjunction with the market crash in 2020.   
	
	\subsection{Literature Review and Empirical Motivation}\label{S1.1}
		There are three branches of literature that the current study engages with. The first is on the methodological side and relates to model selection, where attention to time-variability is crucial. Time-varying functional forms have been addressed in deep learning, for instance, by extending traditional feedforward neural networks	to LSTMs \cite{Hochreiter1997}, which are more suited to financial time series \cite{SirignanoCont}. Time series econometricians refer to this problem as "time-varying parameters" \cite{ACH13, Coulombe21}. The field of change-point detection is also relevant, as it deals with shifts in the underlying distribution of a time series \cite{BCF19, FR19}. Adaptive time series methods are another set of attempts at time series model selection, as \cite{Yang2020, Yang2021} demonstrate the possibility for lag operators and window selection based on the loss induced by the past forecasts. The DMS and AE methods directly relate to the model selection and ensembling in a time series environment: the loss functions and sample evaluation methods account for the usual concerns of stability and interpretability while being engineered specifically for time series. 
		
		The second branch is related to empirical risk minimization in the statistical learning literature. In particular, robust statistics and forecasting theory \cite{granger1999outline, granger2002some, HR2009, FR19} play a role and motivate many potential loss functions for AE and DMS. In addition to that, domain-specific evaluation metrics in a change-point or contaminated environment may differ hugely from traditional metrics such as Mean Squared Error (MSE), and may be more robust instead. To this point, recent literature demonstrates how financial metrics can be useful as a contextual evaluation for various statistical or machine learning models \cite{OC2020, BCR22, Li2021, Sood21}. Discussions on Sharpe Ratio (SR) as a statistical measurement can be further traced to \cite{AL2002, Opdyke07}. Crucially, SR serves not only as a tool for portfolio evaluation in finance, but also as an evaluation metric for financial time series. This motivates DAA to be introduced as a tool for dynamic and contextual evaluation in financial time series. Combining the algorithms practically extends to the concept of automation, such as the proposal of `Automated Statistician' \cite{steinruecken2019a}. In this paper, we extend our method into a potential `Automated Portfolio Manager', where the machine learning aspect of the risk-minimisation process could be translated to user-algorithm interaction in financial practice.
		
		The final branch of literature concerns the empirical findings from the Yield Curve, VIX Curve, and financial markets. There is a large body of existing literature linking the Yield Curve to future economic growth and many other important macroeconomic variables \cite{estrella1996, DePace13}, especially prior to the downturn of the economy. \cite{Yang2020}, for example, finds empirical success in forecasting US GDP Growth using the Yield Curve. Likewise, the VIX Curve has become widely recognised amongst the practitioner community as an important signal for the future state of the stock market \cite{fassas_2012, Fassas_19, whaley2009}. Still, despite numerous studies suggesting the existence of a relationship, a detailed forecasting study has yet to be conducted; neither has a study been done using non-traditional evaluation metrics and statistical learning techniques. As an empirical observation, \autoref{fig0} shows the state of these relationships on 24 Feb 2020, demonstrating that inversions of both curves can signal a downturn in the stock market. We further this analysis and provide a case study on the 2020 market crash.

	\subsection{Key Problems and Contributions}
		We state the following two problems we aim to tackle.
			\begin{itemize}
				\item 	\textbf{Problem 1}:  In time series, like many other types of data, a common problem we face is how to select a set of models or combine them. However, classic techniques such as cross-validation would rely on certain notions of stability or stationarity in the distribution, whereas in time series, abrupt regime switches frequently occur and disrupt the robustness of such a selection. 
				
				The problem is hence: what loss function and sample evaluation method should be used to select the best parametric time series model, and how to ensemble different models over time?
				
				\item 	\textbf{Problem 2}:  For a portfolio manager, a common problem is how to compare and select from competing models and trading strategies subject to investment constraints. With the common evaluation metrics such as MSE, we are restricted to compare predictive models within the same variable and same forecast horizon: at time $t$, let $k_1, k_2 > 0$ while $k_1 \neq k_2$, if we forecast $Y$ at time $t+k_1$ conditional on present ($Y_{t+k_1|t}$), the distribution of itself could be quite different from the distribution of $Y$ at time $t+k_2$ ($Y_{t+k_2|t}$) or another variable $Z$ at another time. 
				
				The problem becomes: how to compare and select from models with different variables and forecasting horizons? Further, how to automate such a process subject to investment constraints and can this outperform the market?
			\end{itemize}

	Our key contributions in response to the above are as follows. We propose two algorithms (DMS and AE) to respond to problem 1, which serve as general procedures to select and combine models in time series. We designed loss functions that are tailored to time series data and forecasting. In particular, they weigh more heavily recent forecasting errors, providing a more up-to-date evaluation. For models with a larger forecasting horizon, we also consider higher-order forecasting information, such as the loss associated with forecasts made on the intervening horizons.
	
	As a response to problem 2, we propose an evaluation procedure (DAA) to select the outperforming forecasts, for the purpose of their induced financial performance --- the use of SR allows that to draw together different assets and compare across different forecast horizons. Similar to the idea of an `automated statistician' in the recent literature, it is tempting to ask whether it is possible to have an `automated portfolio manager' selecting the top models amongst different variables and forecasting horizons --- DAA is a step forward toward that agenda. More empirically, we address problem 2 by using the aforementioned algorithms to select and invest in different assets, with a number of variations such as capital constraints and hedging. The latter also offers an up-to-date demonstration of the usage of VIX as a hedging tool for speculation, which contributes to the broader literature on portfolio management. 
	
	\subsection{Plan of This Paper}
	
	We present the main algorithms in section \ref{S2}, followed by empirical financial results of the algorithms in sections \ref{S3} and \ref{S4}. We also show how the algorithms' outputs can be interpreted by examining the environment of 2020 market crash in section \ref{S5}.   Conclusions and future plans are then made in section \ref{S6}.

\noindent
\begingroup
\removelatexerror% Nullify \@latex@error
\begin{algorithm*}[t]
	\caption{DMS}\label{algALDMS}
	\SetAlgoLined
	\KwIn{Data, desired forecasting index set $T$, and hyperparameters $(\ell, H, \{\Xi_{h,i}\}_{i \in I(h), h \in H}, v)$}
	\KwOut{Forecasts $\{\hat{y}_{t+k|t}(h_t^{DMS})\}_{t \in T }$ with the associated models $\{h_t^{DMS}\}_{t\in T}$}
	\begin{enumerate}
		\item For $t\in T$, repeat:
		\begin{enumerate}
			\item Evaluate $\ell$ given the information required. Then find $h^* \in H $ and $ \Xi^*_{h,i} $ which minimises the loss.
			\item	 Obtain and store $\hat{y}_{t+k|t}(h_t^{DMS}):=\hat{y}_{t+k|t}(h^*, \Xi^*_{h,i} )$ as the forecast
		\end{enumerate}
		%\vspace{-0.25cm}
	\end{enumerate}		
\end{algorithm*}
\begin{algorithm*}[t]	
	\caption{AE}\label{algALEnsemble}
	\SetAlgoLined
	\KwIn{Data, desired forecasting index set $T$, and hyperparameters $(\ell, H, \{\Xi_{h,i}\}_{i \in I(h), h \in H}, v_0, v_1)$}
	\KwOut{Forecasts $\{\hat{y}_{t+k|t}(h_t^{AE})\}_{t \in T }$ with the associated models $\{h_t^{AE}\}_{t\in T}$}
	\begin{enumerate}
		\item Enumerate $\cup\{(h,\Xi_{h,i}) : i \in I(h), h\in H  \}$ to $[M]$. For $t\in T$, repeat:
		%\item 
		\begin{enumerate}
			\item For $\tau \in \{t-v_0+1,..., t\}$, repeat:
			\begin{enumerate}
				\item Evaluate $\ell$ given the information required. Then find $h^* \in H $ and $ \Xi^*_{h,i} $ which minimises the loss.
				\item Allocate a weight of $v_0^{-1}$ to the minimiser.
			\end{enumerate}
			\item Collect the weight $\delta_t$ and align the forecast vector $\hat{y}^M_{t+k|t}$
			\item 
			Obtain and store $\hat{y}_{t+k|t}(h^{AE}_t) = \langle \delta_t, \hat{y}^M_{t+k|t} \rangle$ as the forecast
		\end{enumerate}
	\end{enumerate}		
	
\end{algorithm*}

\endgroup
\section{Proposed Methodology}\label{S2}
	We consider the problem of selecting from a set of parametric time series models, denoted here by $H$. For example, $H$ may contain AR models up to certain lags, and also various models including other explanatory variables. For each model $h \in H$, we have various estimation techniques, e.g. MLE or Yule-Walker Method with various window sizes ($w$) which control the sample size of the parametric estimation. We denote such a technique as $\Xi_{h,i}$, where $i \in I(h)$ serves as an index. 
	
	\subsection{Dynamic Model Selection}
		We first introduce the case of finding $(h^*, \Xi^*_{h,i})$, the optimal model and estimation technique pairing. We refer to this algorithm as Dynamic Model Selection (DMS), and annotate the selected model as $h^{DMS} \in H$. The key here is to select the model at time $t$ and adapt the forecast to the selected one. In particular, we use global loss function  $\ell$ to facilitate the evaluation of the `goodness' of the model, therefore returning $\hat{y}_{t+k|t}(h^{DMS})$. 
		
		Our approach is designed to find the model(s) with the lowest loss, which is closely related to out-of-sample forecasting error. We do so by considering various local loss functions and sample evaluation methods (what can be thought of as the global loss function). Though this may be analogous to empirical risk minimization, which is commonly implemented, we propose a variety of loss functions aiming to achieve a more accurate model selection by robust evaluation on past forecasts. To do this, we evaluate the loss of each model over a recent tranche of data.	Consider a $v$-sized window up to time $t$, $\{y_\tau\}_{\tau=t-v+1}^t$, and a given pair $(h,\Xi_{h,i})$. We have access to $\hat{y}_{\tau|\tau-\tilde{k}}(h,\Xi_{h,i})$ for all $\tau \in \{t-v-k+1, ...,t-1, t\}$ and $\tilde{k} \in [k]$.  Now, define a formulation of $\ell$ as below:
		\begin{equation}
			\ell^{\texttt{single-valued}}(h,\Xi_{h,i}; \lambda,p):= \sum_{\tau=t-v+1}^t \lambda^{t-\tau} |\hat{y}_{\tau|\tau-k} - y_\tau |^p \label{Nsv}
		\end{equation}
		\autoref{Nsv} specifies a global loss function by summing over the local loss functions in a window $v$. The associated hyperparameters are $\lambda \in (0,1]$ and $p \in (0,\infty)$. The larger the $\lambda$, the more focused the loss is towards the recent history, and vice versa. While $v$ is clearly a hyperparameter as well, it is not being investigated in this paper, as $\lambda^{t-\tau}$ discounts the history in a way that controls the effective contribution of the far history to the loss. For general intuition, take $\lambda=1$ and $p \in \{1,2\}$, then \autoref{Nsv} is simply an MAE or MSE over the period. 
		
		We next consider a loss function that is more suited to models with a longer forecast horizon, shown in \autoref{Nmv}, where $\bm{1}_k:=(1,1,...,1,1) \in \mathbb{R}^k$ and $\bm{\hat{y}}_{\tau+k|\tau}:=(\hat{y}_{\tau+k|\tau+k-1}, \hat{y}_{\tau+k|\tau+k-2},...,\hat{y}_{\tau+k|\tau}) \in \mathbb{R}^k$. 
		\begin{equation}
			\ell^{\texttt{multi-valued}}(h,\Xi_{h,i}; \lambda,p):= \sum_{\tau=t-v+1}^t \lambda^{t-\tau} ||\bm{\hat{y}}_{\tau|\tau-k} - y_\tau \bm{1}_k ||_p^p  \label{Nmv}
		\end{equation}
		The novelty behind \autoref{Nmv} is as follows. For $k\geq 2$, the term that differs from \autoref{Nsv} is  $\bm{\hat{y}}_{\tau|\tau-k} - y_\tau \bm{1}_k$. This is a vector containing information in a higher dimension than just the scalar $\hat{y}_{\tau|\tau-k} - y_\tau$. 
		For instance, the loss within \autoref{Nmv} 
		contains $\hat{y}_{\tau|\tau-1}-y_\tau$ due to the vector form and the norm taken --- this is not part of \autoref{Nsv}.
		We note that $\hat{y}_{\tau|\tau-1}$ uses information up to $\tau-1$, which is more recent than the one used by $\hat{y}_{\tau|\tau-k}$, hence facing less noise and more contemporary information on the performance of model and estimation techniques.
		
		Following the above configuration of the loss function, we find the pair which minimises the loss: \begin{equation}\label{Minimisation}
			(h^*, \Xi^*_{h,i} ) := \argmin_{(h,\Xi_{h,i}) \in  \cup\{(h,\Xi_{h,i}) : i \in I(h), h\in H  \}
			} \ell(h, \Xi_{h,i})
		\end{equation}
		DMS then adapts to the selected model for forecasting: $\hat{y}_{t+k|t}(h^{DMS}_t) = \hat{y}_{t+k|t}(h^*, \Xi^*_{h,i} )$. For a general algorithm for computing implementation, see \autoref{algALDMS}.

	\subsection{Adaptive Ensemble}	
		We next introduce the case of ensembling models according to their loss over a recent window of data. We call this Adaptive Ensemble (AE). The ensembling weight takes the form of a convex combination from $H$, in particular, the weight lies in  a simplex $H^{AE} = \Delta(M)$ where $M:= |\cup\{(h,\Xi_{h,i}) : i \in I(h), h\in H  \}|$, which is the total number of estimation techniques across all models. (There are various ways for ensembling models, some more closely related to time series, such as the Particle filters methods \cite{Djuricetal2003}. Here, we propose to form the weights in an additive sub-sampling scheme without Bayesian inference.)
		
		Algorithmically, we first enumerate the set $\cup\{(h,\Xi_{h,i}) : i \in I(h), h\in H  \}$ into $[M]$ and align the associated forecasts to a M dimensional vector $\hat{y}^M_{t+k|t}$. For an element called ensemble weight $\delta_t \in \Delta(M)$, the AE adapts to the selected weight by the inner product: 
		$\hat{y}_{t+k|t}(h^{AE}) = \langle \delta_t, \hat{y}^N_{t+k|t} \rangle$. It remains to introduce the method of determining the weight: we subsample the data points in the window $v$ as follows. Consider a smaller window $v_0$, and for each time $\tau \in \{t-v_0+1,...,t-1,t\}$, look back a window of $v_1$, where $v_1=v-v_0$. Find the best model in the $v_1$ window up to each time $\tau$. That is, for each $\tau$, we proceed with the optimisation as per \autoref{Minimisation}. Allocate an equal weight of $v_0^{-1}$ to the model selected. This then produces $\delta_t \in \Delta(M)$ as desired. See \autoref{algALEnsemble} for a general algorithm.
		
	\subsection{Trading strategies and evaluation metrics}
	\noindent
	\begingroup
	\removelatexerror% Nullify \@latex@error
	\begin{algorithm*}[h]
		\caption{Dynamic Asset Allocation (DAA)}\label{DAA}
		\SetAlgoLined
		\KwIn{Quarters $\{q_j\}_{j=1}^J$, SR of associated strategies $S_j^{k,A}(h)$ for each $j\in [J], k \in [K], h \in H^A, A \in {\mathcal A}$, Capped or Uncapped}
		\KwOut{Selected $N$ strategy for each quarter.}
		\begin{enumerate}
			\item For each quarter $j\in [J]$:
			%	\begin{enumerate}
			\\ If Uncapped:
			\begin{enumerate}
				\item Collect ${\mathcal{S}}_j:=\{S_j^{k,A}(h): k \in [K],
				h \in H^A, A \in {\mathcal A} \}$.  Rank and select the top $N$ candidates from ${\mathcal{S}}_j$. 
				\item Run the selected strategies for the next quarter, then average the holding weights across all selected strategies.
			\end{enumerate}
			Else (Capped): 
			\begin{enumerate}
				\item For each $A \in {\mathcal A} $:
				\begin{enumerate}
					\item  Collect ${\mathcal{S}}^A_j:=\{S_j^{k,A}(h): k \in [K],
					h \in H^A\}$ Rank and select the top $K$ candidates from ${\mathcal{S}}^A_j$.
					\item  Run the selected strategies for the next quarter, then average the holding weights across all selected strategies.
				\end{enumerate}
				\item Average all strategies over all assets.
			\end{enumerate}
			%	\end{enumerate}
			\item Return the selected strategies.
		\end{enumerate}		
	\end{algorithm*}
	\endgroup	
		As was reviewed in section \ref{S1.1}, MSE is unfit for the purpose of evaluating forecasts in the context of financial time series. Here, we consider the following financial evaluation metrics from an induced trading strategy. These metrics form the backbone of the DAA method we propose in section \ref{S2.4}. 
		
		Denote $\{1,2,...,T\}$ as the index of testing data, and as the frequency of data is daily, each index corresponds to one trading day. Recall $k$ is fixed as a forecast horizon. Let $\hat{y}^A_{\tau|\tau-k}$ be the $k$-step ahead forecast made on the return of asset $A$ at time $\tau-k$. At time $t$, we hold $w_t^A$ of asset $A$ as per \autoref{weights_eq0}.
		\begin{equation}\label{weights_eq0}
			w_t^A := \frac{1}{2} +  \frac{1}{2k} \sum_{j=0}^{k-1} (\mathds{1}[\hat{y}^A_{t+k-j|t-j}  >0] - \mathds{1}[\hat{y}^A_{t+k-j|t-j}  <0] )
		\end{equation}
		It is important to observe \autoref{weights_eq0} as an effective restriction to long-only holdings. Alternative trading strategies use long-short signals, which implies solely trading on the direction of the model's prediction. This is a limitation of the current study, since the $\frac{1}{2}$ harmonises the performance by endowing us with a base case of half of the baseline. However, it also eases the debate towards costs of shorting, as shorting an asset may not be always feasible and can be costly, especially for VIX. Certainly, the notion of `short the market' may be achieved by long VIX, and we investigate this further in the form of cross asset strategy in section \ref{S4}.
		
		Let $P^A_t$ be the price of asset $A$ at time $t$. Consider the notion of profit or loss at time t+1 as $\pi_{t+1}^A := w_t^A \times \frac{P_{t+1}^A - P_{t}^A}{P_t^A}$.  Denote $m_\pi^A$ and $s_\pi^A$ as, respectively, the mean and standard deviation of $\{\pi_{\tau}^A\}_{\tau=2}^T$.  The annualised return ($ANR$) then follows the accounting of profits or loss, defined as $ANR = 252 m_\pi^A$
		and the annualised Sharpe Ratio ($SR$) is defined as $SR = \sqrt{252} m_\pi^A (s_\pi^A)^{-1}$.
		Another performance measure for risk is known as Maximum Drawdown (MDD), defined as  $MDD =  \underset{t \in [T]}{\min} \left(  \frac{ \pi_t + 1 }{ \underset{\tau \in [t] }{ \max} (\pi_\tau+1) }  -1 \right)$.
				
		\subsection{Dynamic Asset Allocation}\label{S2.4}

		In time series, we test the predictive models via a `dynamic evaluation' regime, wherein some previous batches of observations would be used for finding the top performers, followed by the next batch of observations for testing. Now, in the context of financial time series, such an evaluation can be attributed to financial metrics rather than MSE. In what follows, we present an algorithm for using SR to decide, on a quarterly basis, what hyperparameter (this includes different algorithms, loss functions, forecast horizons, and assets) to choose. This algorithm is also a step towards an automated portfolio manager, which aims to achieve outperforming financial results while satisfying investment constraints.
		
		To introduce the notation, let $\{1,2,...,K\}$ be the forecast horizons produced, and let ${\mathcal A}$ be the set of all assets. The aim is then to produce an averaged holding of $N:= K \times | {\mathcal A} |$ strategies, from the strategies induced by a collection of models, noted $H^A$ for asset $A \in {\mathcal A}$. (As a note to avoid confusion: the set $H^A$ includes all strategies associated with all hyperparameters and all forecast horizons for asset $A$.)  Now, consider quarter-evaluation dates $\{q_j\}_{j=1}^J$, then for each strategy and asset, we are able to access its historical SR (here, we set this as the recent 252 days, corresponding to one trading year), noted $S_j^{k,A}(h)$. 
		
		\begin{figure}[t]
			\centering
			\includegraphics[width=\linewidth]{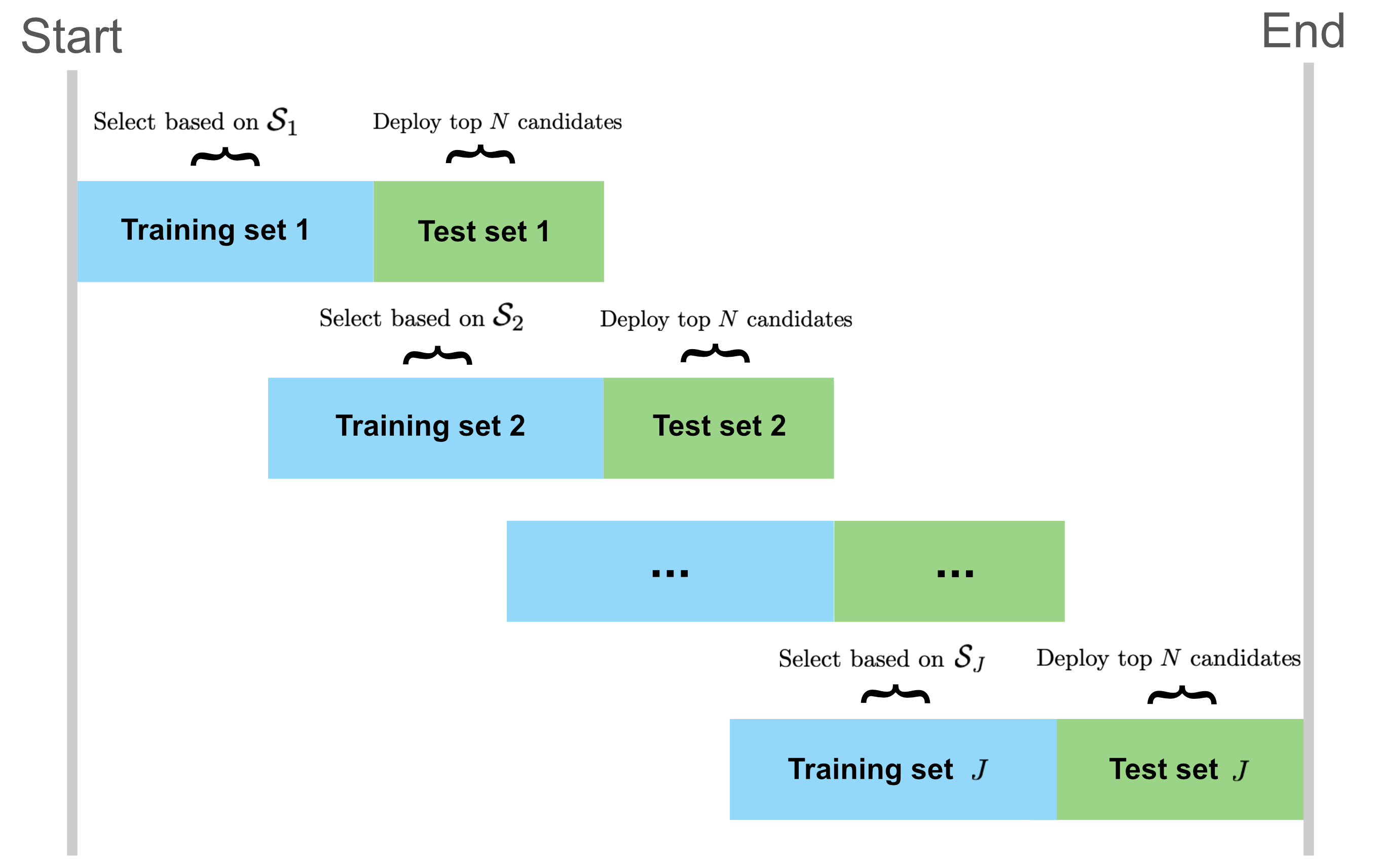}
			\caption{DAA as a rolling process for validation and testing}
			\label{fig:alfinres0}
		\end{figure}
	
		The last input to consider is whether a choice of cap is employed --- this is entirely user-dependent as this relates to the demand from the end-user, in the financial context, the portfolio manager or investor. Choosing the strategy to be capped would mean that the portfolio would be restricted to a certain percentage for each asset's exposure (here, we assume an equal-cap is enforced, so having a capped strategy implies capping the maximum holding of every asset to a weight of $|{\mathcal A}|^{-1}$), else we consider the strategy to be uncapped, where no bounds on the exposure of a certain asset would be enforced. Both strategies have merits and supporting intuitions, being more generally known as dynamically weighted or equally weighted strategies in portfolio finance. As a remark on the case of a single asset, i.e. $| {\mathcal A} | =1$: the DAA, no matter capped or uncapped, reduces to a simpler decision amongst models and forecast horizons.
		
		The full algorithm is provided in algorithm \ref{DAA} and a graphical illustration of the process is provided in \autoref{fig:alfinres0}. Notice that the use of SR has a significant advantage compared to MSE in terms of variable selection: that not only could comparison be made across models, but also across different forecasting variables. This is otherwise not achievable. In essence, at time $t$, if we forecast $Y$ at time $t+k$, the distribution of such variable could be quite different from the distribution of another variable $Z$ at time $t+k$. Now, the use of \autoref{weights_eq0} and associated Sharpe Ratio moderates the values of forecasts, and finally enables a more robust evaluation for model performance, across different variables ($A \in {\mathcal A}$) and forecast horizons ($k\in [K]$).

\section{Financial Results}\label{S3}
	\subsection{Data, Models, and Implementations}
		For the empirical investigation, we use the following data, all on a daily basis: \begin{itemize}
	\item Assets: SP500 index, CBOE VIX index, NASDAQ100 index (NAS100), and DJIA30 index. These are obtained as daily close levels from Yahoo Finance.
	\item Yield Curve (US treasury rates): constant maturity of 1, 3, 6, 12, 24, 36, 60, 84, 120, 240, and 360 months. These are obtained from Federal Reserve Bank of St. Louis' online database.
	\item VIX Curve: VIX Index and 1 to 7 months of maturity futures of VIX contracts are obtained as daily close levels from CBOE.
	\end{itemize}
	The full sample ranges from the start of year 2013 to the end of year 2021. 
	
	As a data processing technique commonly used in the relevant empirical literature, we consider the logged difference of asset price to be return, that is, for a given asset $A$, the $k$-days ahead return is defined as $$
		r_{t:(t+k)}^A:=\log(p_{t+k}^A)-\log(p_t^A)$$where $p_t^A$ denotes the price of asset $A$ at time $t$.
	In the models, we use the notation $y_{t}$ to represent $r_{(t-k):t}^A$. As for yield and VIX curves, we consider the `slopes', noted $s_t$ at time $t$, to be estimated by $$
		p_{t,j} = \alpha_t + s_t m_{t,j} + \varepsilon_{t,j},  \ \ \varepsilon_{t,j} \sim N(0,\sigma_j^2) \ \ \forall j \in [J], t$$where $p_{t,j}$ indicates the price of the future (or interest rate) with maturity $m_{t,j}$. For VIX, $J=8$ and for Yield Curve, $J=11$. This is a common method of estimating the slope of a term structure, as is pointed out in section \ref{S1.1}.
	To populate meaningful models into the model space $H$, we consider the following three classes using the data we have: \begin{itemize}
		\item Class 1: AR(p) models on returns $y_t = \alpha + \sum_{j=1}^p \phi_j y_{t-j} + \varepsilon_t$
		where $p\in \{0,1,...,5\}$. Note $AR(0)$ is a constant model.
		\item  Class 2: lagged linear regression with slope or spread from Yield or VIX curves $y_t = \alpha +  \beta s_{t-k} + \varepsilon_t$
		\item Class 3: lagged linear regression with a pair of short-long rates $y_t = \alpha + \beta_1 \text{short}_{t-k} + \beta_2 \text{long}_{t-k} + \varepsilon_t$
		where short and long refer to short-term and long-term rates: in VIX, short-term refers to maturities up to and including 3 months, with the rest being long-term; and in Yield Curve, short-term refers to maturities up to and including 24 months, with the rest being long-term.
	\end{itemize}
	All models are equipped with 5 different window sizes for estimation: 22, 44, 63, 126, and 252, representing a look-back window of 1, 2, 3, 6, and 12 months respectively. Estimation methods for AR models (those in Class 1) are via Yule Walker, and for explanatory models (those in Class 2 and 3) are via OLS. For DMS and AE implementation, hyperparameters are taken as follows: $K=5$, $v_0=v_1=50$, $v=100$ for choices of maximum forecast steps and window sizes respectively, and $p \in \{1, 1.5, 2\}$, $\lambda \in \{0.8, 0.85, 0.9, 0.95, 0.96, 0.97, 0.98, 0.99, 1\}$ for configurations of the loss functions. 		
	\subsection{Overall Evaluation}
\begin{figure}[h]
			\centering
			
			\includegraphics[width=\linewidth]{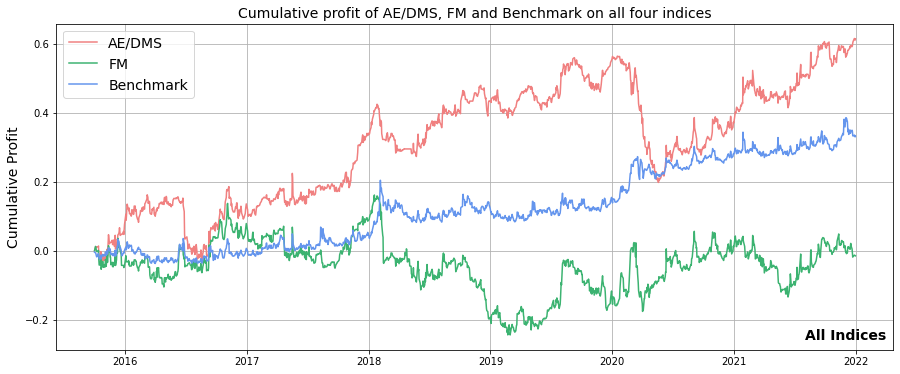}
			\includegraphics[width=\linewidth]{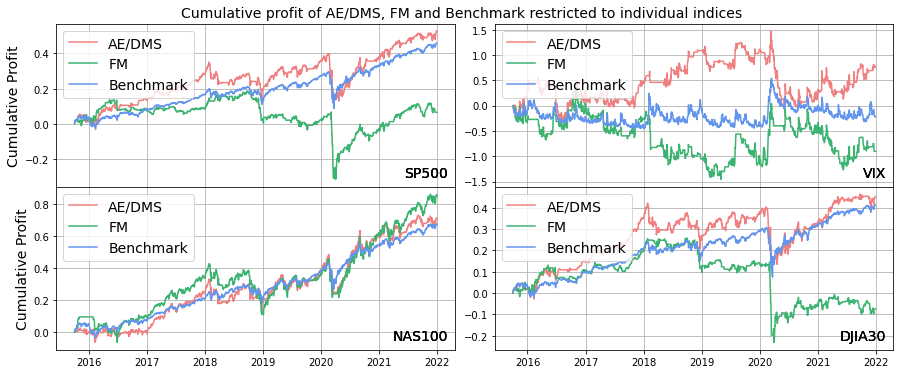}
			%		\vspace{-3mm}
			\caption{Cumulative return of AE/DMS, Fixed models (noted FM in the graph) and Benchmark under trading schemes that consider four indices (upper) and each asset separately (lower)}
			\label{Overall_res}
		\end{figure}

		As an overall evaluation, we implement trading strategies induced by both AE/DMS and fixed models. We  consider trading all four market indices: SP500, VIX, NAS100, and DJIA30. We use the first full year of validation data from Q3 2014, to select the optimal hyperparameters and model specifications. In \autoref{Overall_res}, we provide the cumulative profit associated with both strategies over the testing period, spanning from the start of Q4 2015 to the end of 2021. As a comparison, we also provide a benchmark that holds a constant of 0.5 and weights equally between the four indices.  These results are presented in both individual returns (lower) and averaged (upper). 
		
		The AE/DMS induced forecasts strongly outperform the ones by fixed models, in terms of cumulative profit in the testing period, being up 60\% where the fixed models are approximately break-even. The AE/DMS  also outperforms the equally-weighted benchmark, albeit having a more tumultuous rise. This can also be observed in \autoref{tab:detailed_performance}, where AE/DMS outperforms in terms of ANR and SR, but underperforms the benchmark in terms of MDD. By observing the performance of individual asset's results, we see the AE/DMS outperforms the fixed models in 3 out of the 4 assets, while outperforming the benchmark in all four cases.
		
		\begin{table}[h]
			\centering
			\begin{tabular}{lrrr}
				Strategy &   SR   &   	MDD    &   	ANR     \\
				\hline
				AE/DMS  &   \textbf{0.558} & -23.40\%  & \textbf{9.92\%}  \\
				Fixed models &  -0.012& -34.93\% & -0.24\%\\
				Benchmark&   0.427 &  \textbf{-10.20\%} &      5.36\% \\
			\end{tabular}
			\caption{Performance metrics on all four indices (upper panel of \autoref{Overall_res})}
			\label{tab:detailed_performance}
		\end{table}
	
	\subsection{Towards an Automated Portfolio Manager: Dynamic Asset Allocation}
		We proceed to present the results obtained by selecting different learning-induced strategies over time, as was prescribed in section \ref{S2.4}.  Here, we focus on the three equity indices, and return to discuss how VIX can be used within in a dynamic hedging framework in  section \ref{S4}. As part of the DAA setup, we consider both a capped and an uncapped regime (as detailed in section \ref{S2.4}, the capped regime implies that each index weighting cannot exceed a certain threshold, while uncapped does not impose this restriction). We also include a benchmark, which uses an equal weight in the three equity indices. 
		
\begin{figure}[h]
			\centering
			\includegraphics[width=\linewidth]{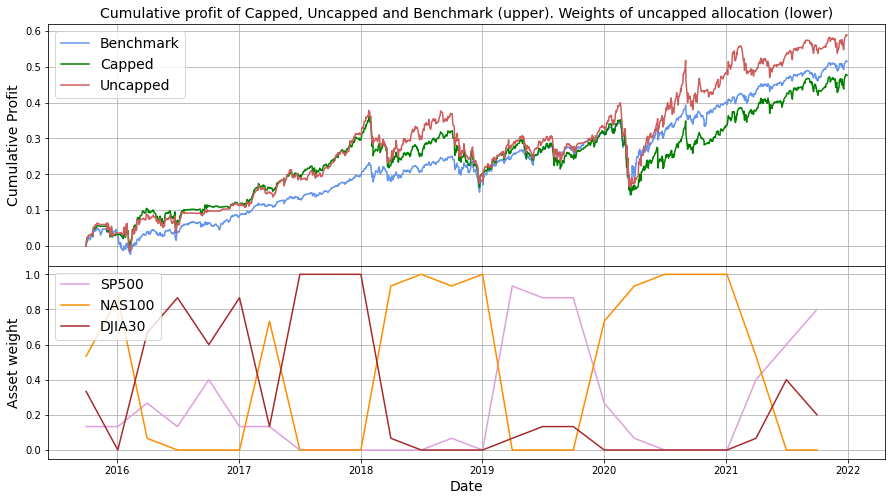}
			\caption{
				Upper: cumulative return over time; Lower: portion allocated to each asset over time in the uncapped regime 
			}
			\label{fig:DAA_res}
		\end{figure}

		The financial results are presented in the upper panel of \autoref{fig:DAA_res}. It is clear that all three strategies perform similarly until 2020, at which point the uncapped DAA outperforms the benchmark in terms of cumulative profit, while the capped DAA underperforms. This may be traced to the weights of the uncapped allocation, shown in the lower panel of \autoref{fig:DAA_res}. In particular, the uncapped allocation keeps a high weighting in the NAS100 during and after the 2020 market crash. Ultimately, the uncapped allocation shows a stronger recovery and greater cumulative profit than both the capped allocation and the benchmark, owing to the strong performance of the underlying NAS100 index. 
		
		The summarised statistics are presented in \autoref{tab:DAA_tab}: both the uncapped and capped allocations obtain a lower Sharpe ratio and MDD than the equally-weighted benchmark. Hence, while varying the weight on each asset (the uncapped DAA does this significantly as per the lower panel of \autoref{fig:DAA_res}) may be more profitable than weighting equally, it does not appear to improve risk-adjusted return or other downside metrics as a whole. This corroborates standard diversification arguments, wherein increasing the number of securities reduces the variance of the portfolio. In \autoref{tab:DAA_tab}, this is presented by a higher ANR and lower MDD, compared to the others.
		\begin{table}[h]
			\centering
			\begin{tabular}{lrrr}
				Strategy& SR  & MDD  &  ANR     \\
				\hline
				Capped	  &   0.639 &   -16.04\%   &    7.68\%  \\
				Uncapped	 & 0.679 &   -16.87\% &    \textbf{9.49\%}   \\
				Benchmark	 & \textbf{0.874} &  \textbf{-14.85\%}&   8.32\% \\
			\end{tabular}
			\caption{Performance metrics for DAA compared with benchmark (\autoref{fig:DAA_res})}
			\label{tab:DAA_tab}
		\end{table}
\section{Extension to Cross-Asset Strategy}\label{S4}	
	\begin{figure}[t]
		\centering
		\includegraphics[width=1\linewidth]{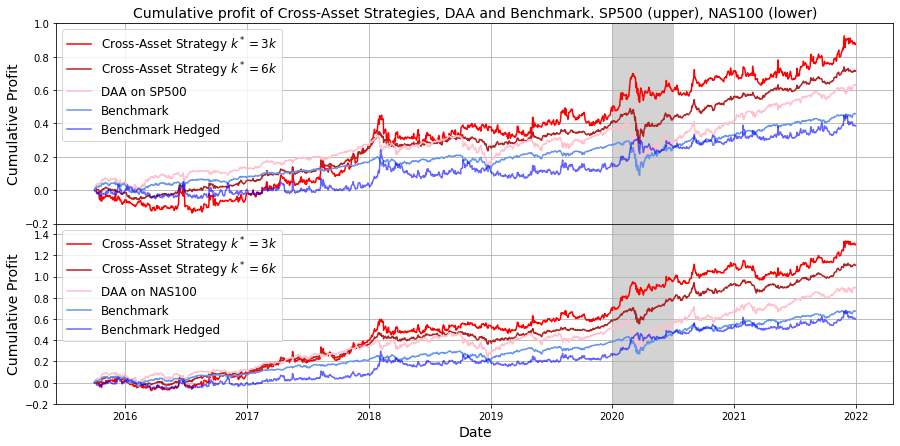}
		\caption{Cross-asset strategies compared to DAA and benchmarks for SP500 (upper) and NAS100 (lower). The shaded area is Q1 and Q2 2020, which are further discussed in \autoref{S5}}
		\label{fig:CAS}
	\end{figure}
	\begin{table}[t]
		\centering
		\begin{tabular}{llccc}
			Asset & Strategy &     SR &     MDD &     ANR \\
			\hline
			SP500 & CAS $k^* = 3k$     &  0.611 & -14.80\% &  \textbf{14.15\%} \\
			& CAS $k^* = 6k$        &  \textbf{0.970} & -14.08\% &  11.51\% \\
			& DAA with no VIX &  0.700 & -23.13\% &  10.17\% \\
			& Benchmark (Long Only)      &  0.807 & -16.00\% &   7.40\% \\
			& Benchmark (Always hedged)  &  0.368 & \textbf{-13.87\%} &  6.23\% \\
			\hline
			NAS100 & CAS $k^* = 3k$        &  0.907 & -12.05\% &  \textbf{21.00\%} \\
			& CAS $k^* = 6k$        &  \textbf{1.303} &  \textbf{-9.29\%} &  17.85\% \\
			& DAA with no VIX &  0.822 & -18.20\% &  14.42\% \\
			& Benchmark (Long Only)      &  1.009 & -11.46\% &  10.88\% \\
			& Benchmark (Always hedged)  &  0.588 & -11.43\% &  9.71\% \\
		\end{tabular}
		\caption{Performance metrics of Cross-asset strategies (CAS), regular DAA and benchmarks}
		\label{tab:detailed_performance_CAS}
	\end{table}
	
	A passive hedge, such as a fixed portion of the VIX, inherently lowers the return of a portfolio. This can be observed in \autoref{Overall_res}, where a constant 1/4 holding of VIX causes the benchmark to underperform in terms of ANR and SR. At the same time, hedging serves an important purpose in limiting an investor's downside. Indeed, \autoref{tab:detailed_performance} also shows that an investor would face a drawdown of at most 10.20\% over the lifetime of this portfolio. This apparent trade-off is consistent with the general conception of hedging in a passive context: during a market upturn, a hedge can cause a so-called 'return-drag',  while in a down-market it can directly offset losses.
	
	This begs the question as to whether we can construct a dynamic hedging strategy to avoid return drag while maintaining downside protection. To this end, we consider a set of cross-asset strategies (CAS) for an investor hedging SP500 or NAS100 using the VIX. We proceed with the introduction of the case of SP500, while the case of NAS100 is exactly the same in methodology.
	
	Here, we hold the SP500 as normal, longing an extra $\frac{1}{2k}$ unit for every stage of rise ($\mathds{1}[\hat{y}^S_{t+k-j|t-j}  >0]$) predicted, as per \autoref{weights_eq1}. 
	\begin{align}\label{weights_eq1}
		w_t^S &:= \frac{1}{2} +  \frac{1}{2k} \sum_{j=0}^{k-1} (\mathds{1}[\hat{y}^S_{t+k-j|t-j}  >0]) \\
		\label{weights_eq2}
		w_t^V &:=  \frac{1}{k^*} \sum_{j=0}^{k-1} (\mathds{1}[\hat{y}^S_{t+k-j|t-j}  >0])
	\end{align}
	However, when we predict a rise in the SP500, we also long the VIX as a method of hedging our added exposure, as per \autoref{weights_eq2}. The extent to which we hedge is itself a parameter, $k^*$, which takes either $3k$ or $6k$ depending on an investor's risk tolerance. $3k$ implies hedging the entire portfolio, fitting for a risk-averse investor, while $6k$ hedges just the speculative long. 
	
	In addition to the desired strategies and long-only benchmark, we also add a benchmark which, on top of holding a constant of $\frac{1}{2}$ of the asset, holds an additional $\frac{1}{6}$ of VIX. We call this the `always hedged' case.
	
	The cumulative return are plotted in \autoref{fig:CAS} for both assets. Detailed performance metrics are presented in \autoref{tab:detailed_performance_CAS}. CAS achieves a reduction in MDD while still maintaining a strong ANR and SR. The CAS using $k^* = 6k$ in NAS100 outperforms the benchmark on every metric, striking the desired balance between return and downside protection. Clearly, using DAA to dynamically hedge the exposure gives an improvement of risk-adjusted return to the portfolio: CAS tactically holds VIX and maintain a competent risk exposure (as demonstrated by \autoref{fig:CAS} and MDD in \autoref{tab:detailed_performance_CAS}) while achieving higher return (ANR as high as 21\% could be achieved), hence the high SR at the end, with the NAS100 case achieving a SR of 1.303.

\section{Case Study \& Interpretation: 2020 Market Crash}\label{S5}
	In this section, we demonstrate the usage of the algorithms to interpret historical events. The 2020 market crash induced huge volatility and contaminated data, which makes itself as a challenging interpretation task. In forecasting analytics, we select the hyperparameter ex-post as we aim for analysing the composition of the forecasts from DMS and AE, whereas in the financial performance part, we select hyperparameter ex-ante using the automated DAA decision (because DAA is using pre-Q1 2020 information for Q1 2020 and pre-Q2 2020 information for Q2 2020).
	\begin{figure}[h]
	\centering
	\includegraphics[width=\linewidth]{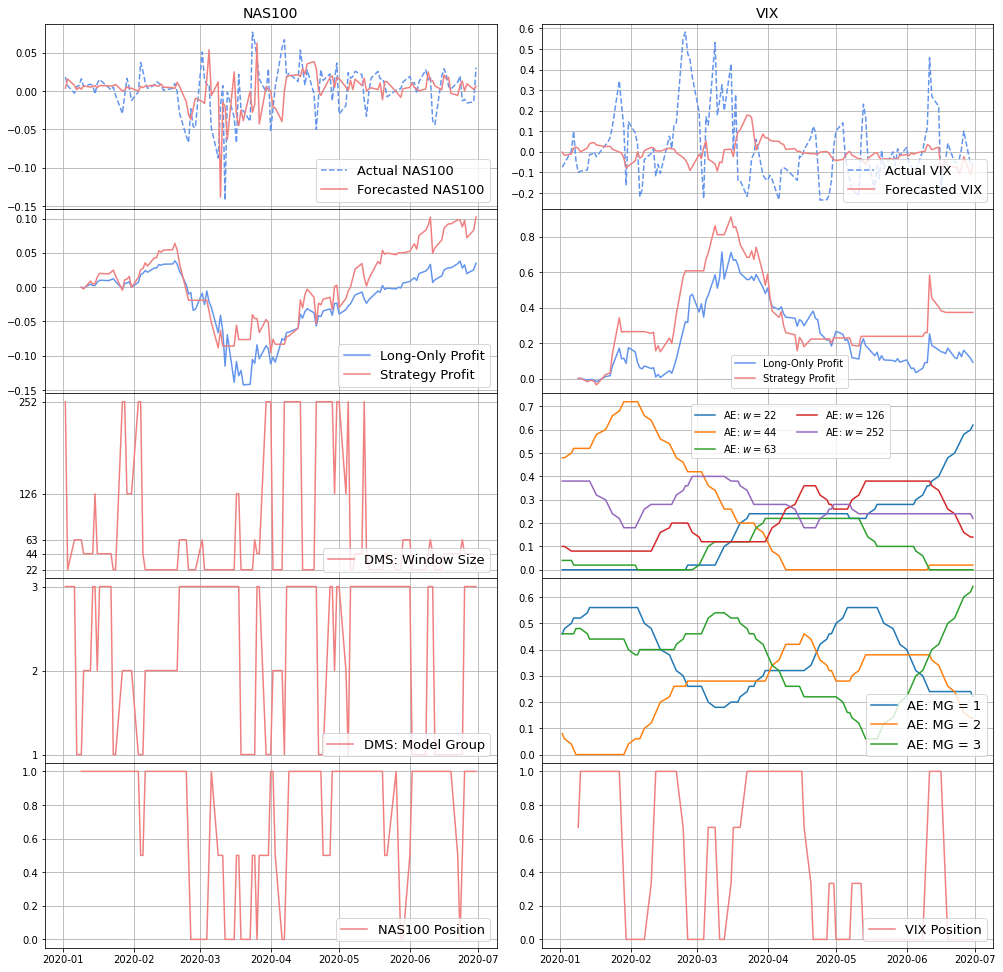}
	\caption{
		Interpretation panel for selected specifications (left: NAS100 with $k=2$ and DMS, right: VIX with $k=3$ and AE). The top row reports the actual returns (blue) versus forecasts (red), the second row reports the induced cumulative profit (red) versus long only (blue), the third row reports the choice of window size (DMS) or weights on the window size (AE), the fourth row reports the choice (DMS) or weights (AE) on model group, and the fifth row reports the induced holding weights from the strategies
	}
	\label{fig:2020anal}
\end{figure}
	\subsection{Forecasting Analytics}  
	We draw attention to \autoref{fig:2020anal}. On the left, we follow the development of NAS100 stock returns with $k= 2$, in which the market crashed and re-bounded in March 2020. The model shown here is a DMS model with $p=0.85, \lambda = 1.5$ and multi-valued norm. On the right, we follow the returns of the VIX index, also with $k = 3$. For this, the model is an AE model with $p=1, \lambda =1$ and single-valued norm (this is exactly the formulation of a rolling MAE). On the lower panels, we show the specifications of the underlying chosen model(s). For NAS100, DMS has a preference for model group 3 during the height of the crash. Similarly for VIX, the AE model  maintains a high weight in model group 3 during this period. Both models favour incorporating explanatory variables from the VIX and Yield curves during this period, which aligns with the perception of the curves as leading market timing indicators.
	
	For window size, the smallest window ($w = 22$) is chosen consistently by the DMS model throughout the most volatile period. This supports the argument that, during a major regime switch, data far in the past can be less relevant to the current evaluation. Hence models incorporating small samples of contemporary data can perform better than those considering large but contaminated samples. When the series stabilises, larger windows ($w$ = 126 or 252) are preferred. Here, larger window sizes ensure greater stability in parameter estimation, since the time series at this point is broadly reflective of the history. This underscores the problem with standard cross validation when being applied to financial time series, as it assumes that the time series distribution would remain unchanged between periods. DMS gets around this problem by providing time-varying cross validation. This allows DMS, for instance, to have a general preference for large window sizes but also to select models with smaller window sizes when larger ones are contaminated.

	\subsection{Financial Performance of Cross Asset Strategies}
	
\begin{figure}[t]
		\centering
		\includegraphics[width=\linewidth]{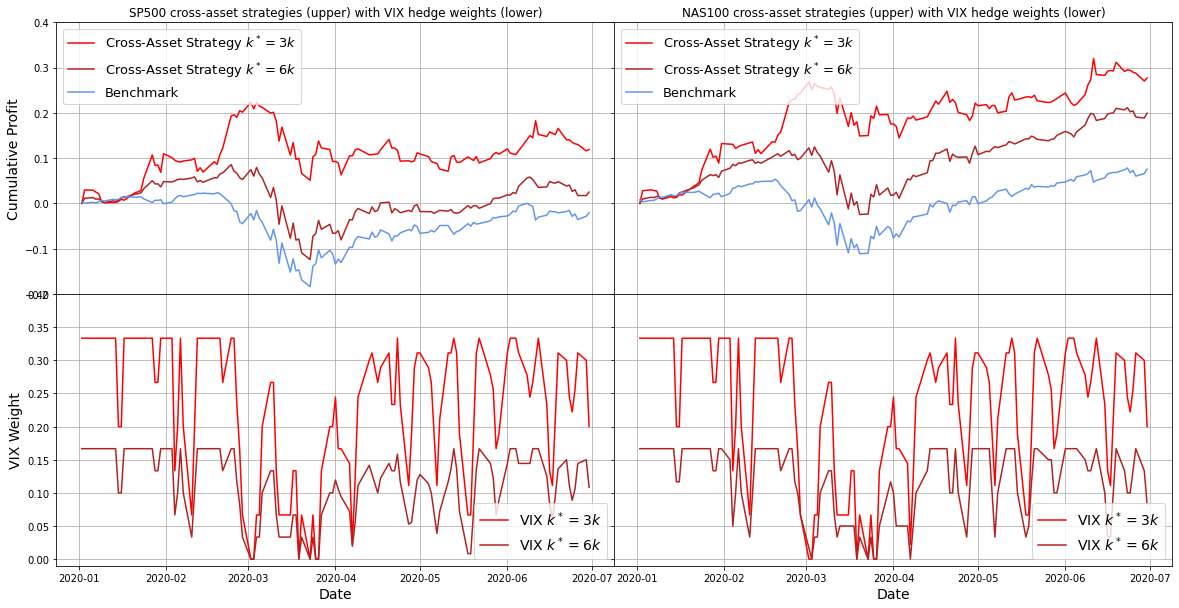}
		\caption{
			Cross-asset strategies compared to benchmark for SP500 (left) and NAS100 (right). The upper plots cumulative return and the lower plots the weights of VIX holdings
		}
		\label{fig:2020CAS}
	\end{figure}

	For financial performance of DAA, we zoom in the results we had on CAS in \autoref{S4}. As plotted in \autoref{fig:2020CAS}, a high return can be observed in the strategies, while the drawdown is maintained to be more robust as its consistent VIX holdings (shown at the bottom) serve as a good hedge. This, together with \autoref{S4}, proves the empirical value of our methods, that they outperform the market in both risks and return, even at the market crash.
	
\section{Conclusion and Future Research Agenda}\label{S6}
	We have proposed a generalised framework and algorithms for time series model selection, ensemble, and financial evaluation. The DMS and AE methods require a specification of a model set in terms of functional forms and training methods for parametric estimation and forecasting. Once equipped, they provide time-varying selection and leverage higher-order time series information than typical cross-validation procedures. 
	
	While various forms of loss functions and hyperparameters were studied in this paper, the algorithm leaves many choices open. The local loss function, for instance, could inherit from robust statistics such as Huber loss; and the global loss could include further techniques of change point detections \cite{BCF19, FR19}. Alternative specifications of hyperparameters ($p$ and $\lambda$) may also further improve the sample evaluation, as was encouraged by \cite{granger1999outline, granger2002some, ACH13}.
	
	Further, we present DAA, which can automatically select from competing strategies. This framework has an edge against typical evaluation metrics, such as MSE, which are incomparable for different choices of the forecasting target. We show that the framework is a step towards an automated portfolio manager, since it can achieve strong returns while obeying typical investment constraints and being interpretable by the end-user. In addition, we extend DAA to cross-asset strategies, where we dynamically hedge against ordinary strategies using VIX. These obtain outstanding financial performance. 
	
	We notice that the trading of VIX could be ambiguous: it is more common to trade the close-to-expiration VIX contracts instead of the index, as the latter incurs continuous adjustment costs due to its original formulation \cite{whaley2009}. A future agenda would be to use these algorithms and seek optimal financial decisions such as the roll-over of contracts.
	
	Lastly, interpretability behind the algorithms can be achieved, as the components of the forecasts can always be retrieved and such statistics can be used for empirical investigations and understandings of the algorithm. We have demonstrated an example by interpreting the model choices during the 2020 market crash, alongside the robust financial results.
	
	From a basic set of parametric models and forecasts, DMS and AE respectively enable model selection and combination over time. DAA further selects the hyperparameters and makes decisions across assets and forecasting horizons of all the models. These all-in-three algorithms enable further extension towards automation in financial time series and machine learning.

\bibliographystyle{ACM-Reference-Format}
\bibliography{bibliography}

\end{document}